\RequirePackage[hyphens]{url}
\documentclass[12pt]{article}
\usepackage{pdflscape}
\usepackage{tikz,natbib,xfrac,amsmath,amsthm,booktabs,subfigure,anysize,hyperref,graphicx,float,tabularx,multirow}
\usepackage[margin=1in]{geometry}
\usepackage{xcolor}
\usepackage{bm}
\definecolor{dark-red}{rgb}{0.4,0.15,0.15}
\definecolor{dark-blue}{rgb}{0.15,0.15,0.4}
\definecolor{medium-blue}{rgb}{0,0,0.5}
\definecolor{ChadBlue}{rgb}{.1,.1,.5}
\definecolor{ChadDarkBlue}{rgb}{.1,0,.2}
\definecolor{ChadBlue}{rgb}{.1,.1,.5}
\definecolor{ChadRoyal}{rgb}{.2,.2,.8}
\definecolor{ChadGreen}{rgb}{0,.4,0}    % Dark Green
\definecolor{ChadRed}{rgb}{.5,0,.5}  % purple
\hypersetup{
    colorlinks, linkcolor={ChadRed},
    citecolor={ChadBlue}, urlcolor={medium-blue}
}

\usepackage[font=small,labelfont=bf,justification=justified,singlelinecheck=false]{caption}
\usepackage{amssymb}
\usepackage{amsfonts}
\usepackage{amsmath}
\usepackage{setspace}  %using this package conflits with footnote hyperlink jump. Using this package makes the footnote links inactive.
\usepackage[all]{hypcap}        % needed to help hyperlinks direct correctly;

\graphicspath{{figures/}}
\usepackage[english]{babel}
\usepackage{longtable}
\usepackage[singlelinecheck=off]{caption}
\usepackage[singlelinecheck=false]{caption}
\usepackage{array}
\usepackage{float}

\usepackage [autostyle, english = american]{csquotes}
\MakeOuterQuote{"}
\usepackage{chngcntr}
\usepackage{rotating, graphicx}
\usepackage{epstopdf}
\usepackage{threeparttable}
\usepackage{sectsty}
\sectionfont{\large}

\usepackage{longtable}
\usepackage{booktabs}
\usepackage{array}
\newcolumntype{R}{>{\raggedleft\arraybackslash}p{2.5cm}}
\newcolumntype{Q}{>{\raggedright\arraybackslash}p{14cm}}
\newcolumntype{U}{>{\raggedright\arraybackslash}p{11cm}}

\usepackage{tabu}
\usepackage{dcolumn}
\usepackage{placeins}
\usepackage[T1]{fontenc}

\title{Comments are welcome\thanks{I thank Juan de Lucio, Benedikt Heid and Francisco Requena for very valuable comments and suggestions. I also thank the feedback from participants in the research seminar at the University of Valencia. I gratefully acknowledge financial support from the Spanish Ministry of Science, Innovation and Universities (RTI2018-100899-B-I00, co-financed with FEDER) and the Basque Government Department of Education, Language policy, and Culture (IT885-16).}}

\author{\large {Asier Minondo}\thanks{Deusto Business School, University of Deusto, Camino de Mundaiz 50, 20012 Donostia - San Sebasti\'{a}n (Spain). Email: \href{mailto:aminondo@deusto.es}{aminondo@deusto.es}}}

\date{ \today \\  }
\usepackage[normalem]{ulem}

\begin{document}

\maketitle

\begin{abstract}
Scholars present their new research at seminars and conferences, and send drafts to peers, hoping to receive comments and suggestions that will improve the quality of their work. Using a dataset of papers published in economics journals, this article measures how much peers' individual and collective comments improve the quality of research. Controlling for the quality of the research idea and author, I find that a one standard deviation increase in the number of peers' individual and collective comments increases the quality of the journal in which the research is published by 47\%.  
\end{abstract}

\begin{flushleft}
\end{flushleft}

\newpage %\setcounter{page}{1}
\onehalfspacing

\section{Introduction}
Scientific progress is fueled by new ideas. During the process of transforming new ideas into research outputs scholars rely on their peers to identify weaknesses in their work, and to find alternative models, methodologies and databases that can improve the quality of their research. Considering the time scholars devote to present draft versions of their papers at conferences and research seminars, and discuss ideas with colleagues, it is reasonable to expect that peers' comments and suggestions should improve the quality of research. However, despite its alleged importance, no study has quantified this contribution yet. The goal of this paper is to fill this gap.

I build a dataset of papers published in economics journals. Based on the acknowledgment section of the paper, I record all the scholars that gave comments on the paper, and the seminars and conferences at which the paper was presented. To obtain unbiased estimates on how these individual comments, seminars, and conferences contributed to the quality of the paper, I control for the quality of the author and research idea. First, \cite{minondo2020presents} shows that high-quality scholars are more likely to be invited to present their work at a research seminar. It also reasonable to expect that papers written by high-quality scholars are more likely to be accepted at conferences. Furthermore, high-quality scholars may receive more comments on their work because they have more opportunities to interact with other scholars at seminars and conferences, or because their work is more likely to be followed. Second, it seems reasonable to expect that scholars will choose their most promising project when deciding what paper they will present at a research seminar and what draft they will send to a colleague. 

To control for the quality of the author, I use the quality of the institution she is affiliated with. In some specifications, I also use author fixed effects. To control for the quality of the research idea, I use a feature of the job placement process of PhD candidates in economics. During their last academic year, future PhD graduates in economics select a project, among their contemporaneous research ideas, as their job market paper. This paper is the tool PhD candidates use to show their research skills to potential employers. Since PhD candidates want to maximize job offers, they select as job market paper their highest quality project. Thus, the fact that a paper was selected as job market paper provides a signal for the initial quality of a research project. I retrieved information from 2067 PhD candidates in economics, from the top US economics departments, that entered the labor market between 2000 and 2018. When the PhD candidate enters the job market, I identify her job market paper and the additional projects she could also have selected as her job market paper. I follow the job market paper and additional projects until they are published. These publications constitute my estimation sample.  

Using the job market status of a paper to control for the quality of the research idea, and author's fixed effect, I find that a one standard deviation increase in the number of individual comments in a paper that received the average number of individual comments, increases the impact factor of the journal in which the paper is published by 16\%. A one standard deviation increase in the number of research seminars in a paper that was presented at the average number of seminars increases the impact factor of the journal by 31\%. Presenting the paper at conferences has no impact on the quality of the journal in which the paper is published, once I control for the number of individual comments and seminars. I find that comments given by high-quality scholars have a larger positive impact on the quality of a paper than comments received from non-top scholars. Likewise, presenting the paper at a top economics department has a larger positive impact on the quality of the journal in which the paper is published than presenting the paper in a non-top economics department. Receiving comments from other scholars and presenting at research seminars have similar effects on theoretical and empirical papers.

This paper is related to the literature exploring how knowledge is produced \citep{stephan2010economicsofscience,fortunato2018scienceofscience} and, in particular, how peers contribute to that process. \cite{azoulay2010extinction,waldinger2012nazigermany,borjas2015whichpeersmatter,agrawal2017howstarsmatter,jaravel2018teamspecific,bosquet2019senders} analyzed how the premature death, migration, or arrival of scientists affect collaborators' and other peers' productivity. My paper contributes to this literature by analyzing another channel by which peers' can affect the quality of a scholar's output: the individual and collective feedback on ongoing research projects. Our finding that peers' feedback has a large positive effect on the quality of research is in line with \cite{oettl2012helpfulness} who found that a scholar's output quality decreases after the death of a co-author if the co-author was helpful to other colleagues.

My analysis is also linked with studies that have analyzed how conferences and meetings contribute to the flow of ideas, to increase the probability of publication, and to enhance the visibility of a paper. \cite{iaria2018frontier} found that the ban on Central scientists from participating at international conferences during and after World War I was associated with a drop in citations between Allied and Central scientists. Using data from the Joint Mathematics Meetings between 1990 and 2009, \cite{head2019geography} showed that a mathematician is more likely to cite the work of another mathematician if they coincided in the same conference. This probability increases if the two scholars coincided in the session in which the cited paper was presented. Using data from a major political science conference that was canceled in 2012, \cite{deleon2018conferences} concluded that the probability that a paper is cited increases by five percentage points over a period of four years if it was presented at the conference. \cite{gorodnichenko2019conferences} found that presenting a paper at major conferences in economics increases the probability of publishing it in a high-quality journal and enhances its visibility. I find that presenting a paper at a leading economic conference is associated with publishing it at a high-quality journal. However, this positive association becomes statistically insignificant once I control for the number of individual comments received by a paper, and the number of research seminars at which it was presented. This paper is close to \cite{brown2005circulating}, who analyzed whether presentations at research seminars, conferences, and comments received from colleagues increase the likelihood that a paper receives a revise and resubmit decision at an accounting journal; and whether individual and collective comments increase the number of citations received by papers published in three leading accounting journals. He finds that presenting at research seminars is the only variable that is positively correlated with receiving an invitation to revise and resubmit, and the number of citations received by a paper. I add to this paper analyzing whether the number of individual and collective comments increase the quality of the journal in which a paper is published. Furthermore, I control for the quality of the research idea and author fixed effects, and explore whether some comments and presentations have a higher impact than others.

The remainder of the paper is organized as follows. Section~\ref{sec:data} describes the dataset and presents some summary statistics. Section~\ref{sec:regressions} discusses the results of the regression analyses, and Section~\ref{sec:conclusions} concludes.

\section{Data}
\label{sec:data}
The sample is composed by PhD candidates from the top 41 US economics departments that entered the labor market between 2000 and 2018. To identify the top US economics departments I use the ranking elaborated by Ideas.\footnote{I use the 10-year ranking of US economics departments published in June 2019. The latest ranking is available at \url{https://ideas.repec.org/top/top.usecondept.html}} Every year, during the fall term, economics departments announce their job market candidates. From the department's web page, I recorded each PhD candidate's job market paper and the projects that she could also have selected as job market paper. These were projects whose sole author was the PhD candidate, or were written with other PhD students. I excluded the papers co-authored with scholars that already had a PhD.\footnote{I included a paper written with a senior scholar if the job market paper was written with the same senior scholar.} I followed the job market paper and papers that could also have been selected as a job market paper until they were published.

Based on the acknowledgment section, I retrieved the information on the number of research seminars and conferences in which the paper was presented, and the scholars that provided comments on the paper.\footnote{I did not include the editors of the journals in the list of scholars that provided comments. I also excluded the acknowledgments for research assistance, sharing data, or facilitating access to data.} Table~\ref{tab:phd_programs} in the Appendix reports the economics departments and the PhD candidate cohorts included in the sample. It also reports, for each PhD program, the number of graduates from which I could retrieve information, and the number of potential job market paper projects that became journal articles. There are differences in the number of PhD candidate cohorts included in the sample across US economics departments. Those differences are explained by the possibility of accessing the information of "old" cohorts. Economics departments provide information about the PhD candidates that enter the labor market in the current year. Few departments also provide links to previous years' job market candidates. To retrieve information for older cohorts, I used the Internet Archive Library (\url{https://archive.org/about/}). In some cases, the library has a fairly complete record of the different versions of the web site over time. However, in many cases, the information is scant, or there is no copy archived. This explains why I could retrieve information for "very old" PhD candidates (i.e., 2000) for some economics departments (e.g., UC Berkeley or MIT), whereas I could only retrieve information about the most recent cohort for others (e.g., Ohio State).\footnote{There is no correlation between non-archived web sites and the quality of the economics departments.}

I measure the quality of a paper with the Scimago Journal Ranking (SJR) of the journal in which it was published.\footnote{This ranking is built using the average number of weighted citations received in the selected year by the documents published in the journal the three previous years.} Similar to \cite{smeets2006newgraduates}, I measure the quality of a PhD candidate by the quality of her placement after graduation.\footnote{If an author reports more than one affiliation I select her latest academic affiliation.} To measure the quality of the placement, I use the worldwide economics institutions ranking elaborated by Ideas.\footnote{I use the 10-year ranking of institutions published in May 2019. The latest ranking is available at \url{https://ideas.repec.org/top/top.inst.all10.html}. The Ideas ranking provides specific scores for the top 5 institutions (494 institutions). For each percentile between 6 and 10, it lists, randomly, the institutions located at that percentile. To provide a score for institutions located between the 6th and the 10th percentile, I ran a regression with the institutions that have a specific score. The dependent variable is the score (in logs) and the independent variables the percentile in which the institution is located (in logs) and a constant. I use the estimated coefficients to calculate a score for percentiles 6, 7, 8, 9 and 10. If an institution is not at the top 10, I assign it the score of an institution located at the 55th percentile.} If a paper has multiple authors I add up the quality of individual authors. I compute the individual comments received by a paper counting the scholars that are listed in the acknowledgments section of the paper. I also compute the number of comments given by top 10 scholars.\footnote{I use the Ideas' author ranking for November 2019. The most recent ranking is available at \url{https://ideas.repec.org/top/top.person.all.html}} I count the seminars and conferences at which the paper was presented. I also count the seminars given at top 10 institutions and, following \cite{gorodnichenko2019conferences}, the presentations at the three major economic conferences: the American Economic Association (AEA), the European Economic Association (EEA), and the Royal Economic Society (RES).\footnote{To identify the top 10 institutions, I use the ranking built by Ideas mentioned above.} 

Table~\ref{tab:data_collection} in the Appendix provides information about the construction and characteristics of the estimation sample. I retrieved information from 2067 PhD candidates that entered the job market between 2000 and 2018. These job market candidates were working on 5118 projects that could have been selected as job market papers. Among those projects, 2070 were selected as job market papers.\footnote{Note that the number of job market papers is larger than the number of job market candidates, since some PhD students have more than one job market paper.} By December 2019, 551 of the job market candidates (27\%) had published their job market paper or another paper they could also have selected as job market paper in a journal included in the SJR. This percentage is in line with the results of previous studies that highlighted the low "publication productivity" of PhD graduates \citep{conley2014productivityphds}.\footnote{Our percentage is even lower than the 40\% figure reported by \cite{conley2014productivityphds}, due to the larger presence of recently graduated students in our sample, whose papers may be still waiting a editorial decision.} A total of 806 out of 5118 potential papers, 16\%, had been published by December 2019. 47\% of these publications were job market papers. This percentage is larger than the share of job market papers among potential projects (40\%). 18\% of publications had more than one author, and 12\% were published in a top 5 economics journal.\footnote{American Economic Review, Econometrica, Journal of Political Economy, Quarterly Journal of Economics, and Review of Economic Studies.} I have information on the number of individual comments for all papers in the sample. However, there are some articles that use formulas such as "we acknowledge \emph{numerous} seminar participants", "\emph{several} audiences", or "seminar and conference participants". Since the number of seminars could not be computed for these publications, the main estimation sample drops from 806 to 685 articles.

\begin{table}[tbp] \centering
\newcolumntype{C}{>{\centering\arraybackslash}X}

\caption{Summary statistics of the estimation sample}
\label{tab:summary}
\begin{tabularx}{11cm}{l r r r r r}

\toprule
{}&{Median}&{Mean}&{SD}&{Min}&{Max} \tabularnewline
\midrule\addlinespace[1.5ex]
All publications&&&&& \tabularnewline
Individual comments&9&10&7&0&49 \tabularnewline
Seminars&1&3&5&0&23 \tabularnewline
Conferences&0&1&1&0&15 \tabularnewline
&&&&& \tabularnewline
Job market papers&&&&& \tabularnewline
Individual comments&12&12&8&0&49 \tabularnewline
Seminars&4&6&6&0&23 \tabularnewline
Conferences&0&1&2&0&15 \tabularnewline
&&&&& \tabularnewline
Others&&&&& \tabularnewline
Individual comments&7&8&6&0&32 \tabularnewline
Seminars&1&1&2&0&21 \tabularnewline
Conferences&0&1&1&0&6 \tabularnewline
\bottomrule \addlinespace[1.5ex]

\end{tabularx}
\end{table}

Table~\ref{tab:summary} reports some summary statistics on the individual and collective comments received by a publication. It provides statistics for all publications, published job market papers, and other publications. The median publication received 9 individual comments. The distribution is not skewed: the average is 10 and the standard deviation is 7. The minimum number of comments received by a publication is zero, whereas the maximum is 49. There are 61 publications, out of 685, with no individual comments. The median publication was presented at 1 seminar only. The maximum number of seminars at which a publication was presented was 23. There are 263 publications, out of 685, that were not presented at any seminar. Note that the distribution of seminars per publication is skewed, since the average number of presentations is much larger than the median. Finally, the median publication was not presented at any conference. The average and the standard deviation is 1. There is a paper that was presented at 15 different conferences, whereas 411 publications, out of 685, were not presented at any conference. Job market papers received more individual comments, and were presented at more research seminars than non job market papers. Specifically, the median job market paper received 5 more individual comments, and was presented at 3 more seminars than a non job market paper.

\begin{figure}[t!]
	\begin{center}
		\caption{Scatter diagrams}
		\label{fig:scatters}
		\begin{tabular}{c}
			\small A. Journal quality vs. Number of personal comments\\
			\includegraphics[height=3.5in]{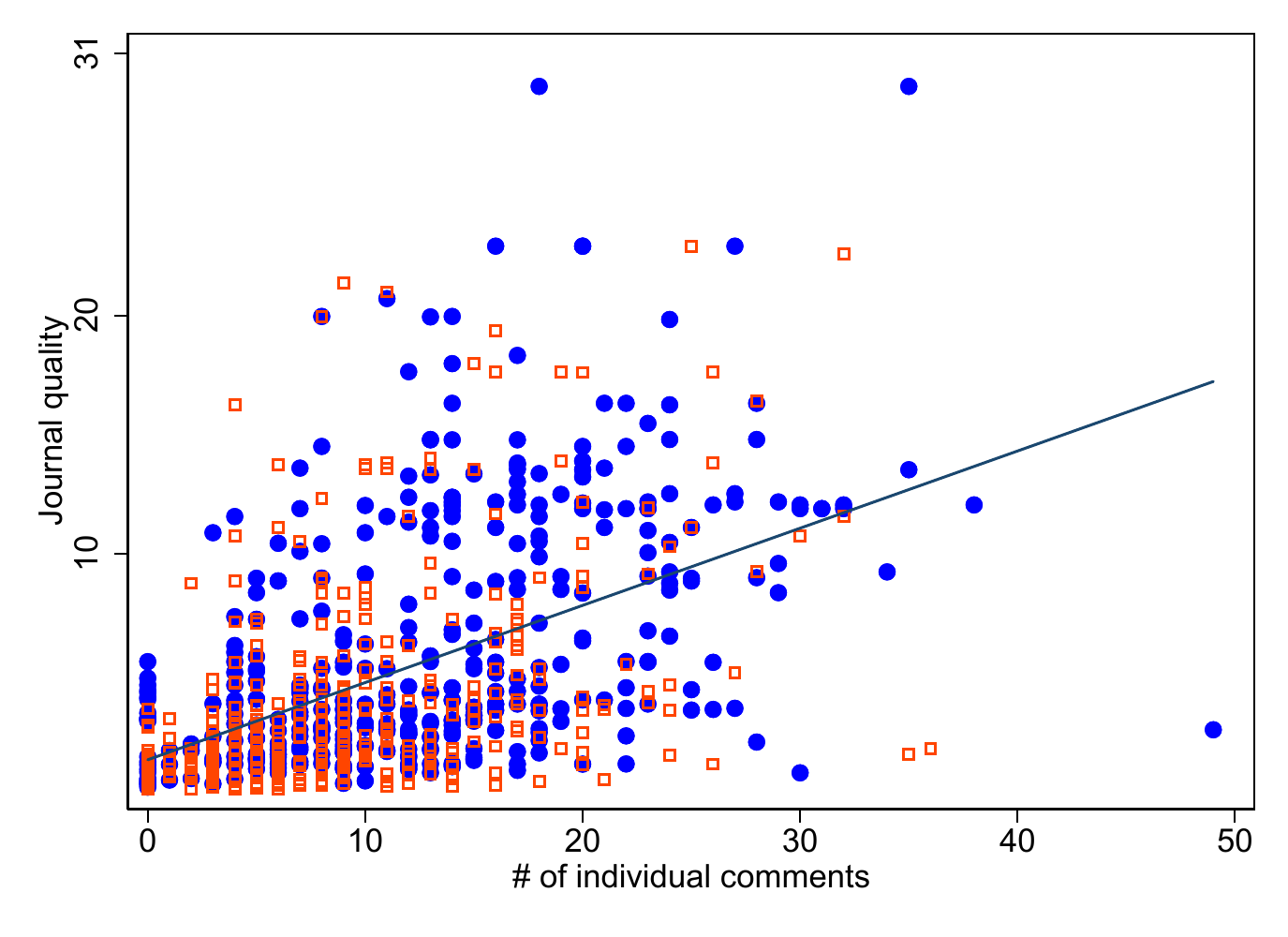}\\
			%	\vspace*{1cm}
			\small B. Journal quality vs. Number of presentations\\
			\includegraphics[height=3.5in]{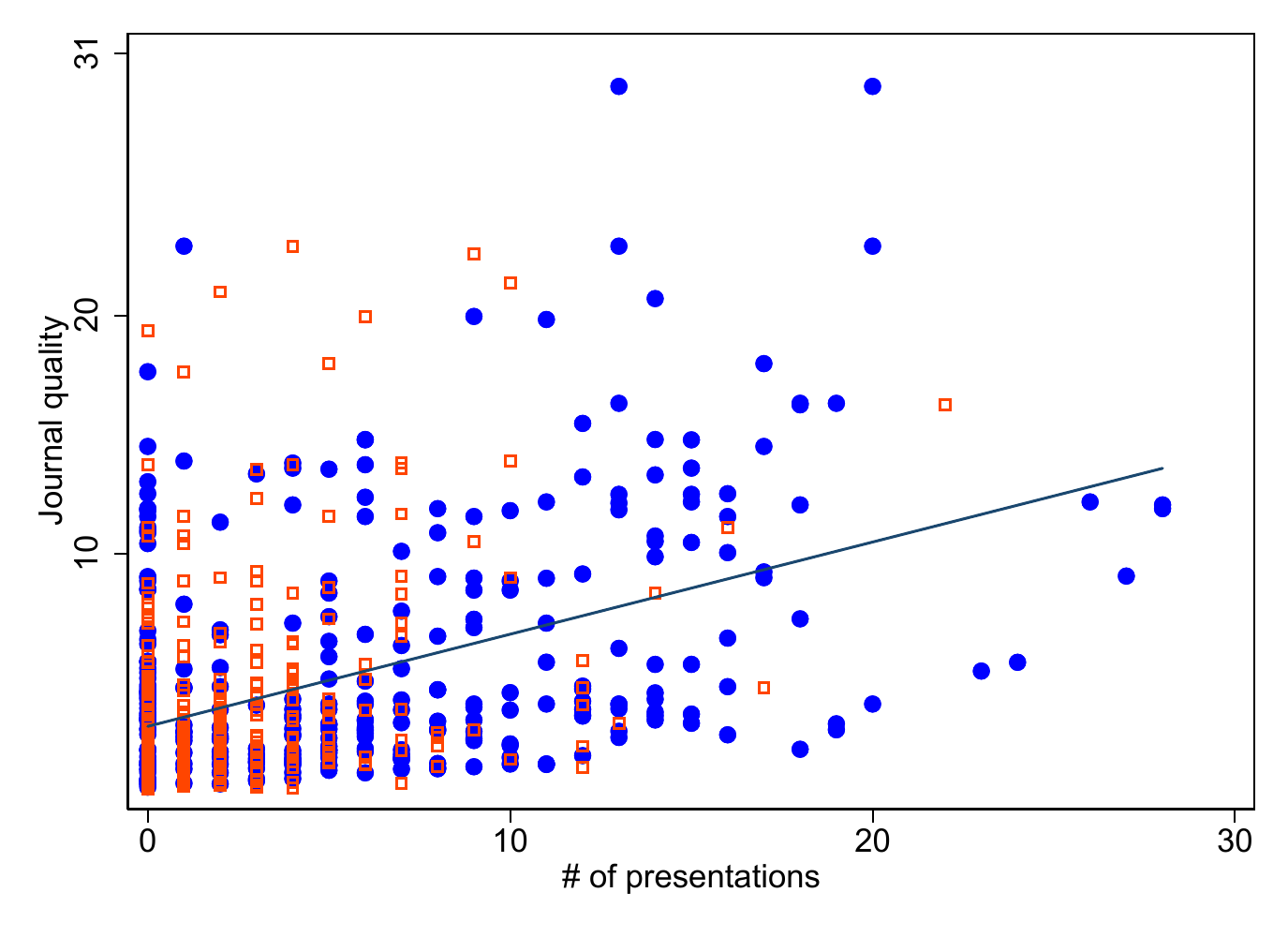}\\
		\end{tabular}
	\end{center}
	\footnotesize{Note: The quality of the journal is measured by the Scimago Journal Ranking. Presentations is the sum of research seminars and conferences at which a paper was presented.}
\end{figure}

Panel A of Figure~\ref{fig:scatters} plots a scatter diagram of the relationship between the number of individual comments received by a paper and the quality of the journal in which it was published. Job market papers are identified by blue dots and non job market papers by red hollow squares. There is a positive correlation between the number of individual comments received by a paper and quality of the journal in which it was published. In Panel B, I add the seminars and conferences in which a paper was presented, and plot a scatter diagram for the relationship between the number of times a paper was presented and quality of the journal in which it was published. There is also a positive correlation between the number of presentations and quality of the journal in which the paper was presented. 

These scatter diagrams suggest that peers' individual and collective comments improve the quality of a paper. However, these correlations may be capturing the positive association between the quality of the scholar and number of individual and collective comments from peers; or the quality of the research idea and comments received from peers. In the next section, I explore the contribution of individual and collective comments to the final quality of a paper once I control for the quality of the author and research idea.

\section{Regression results}
\label{sec:regressions}

To estimate the contribution of peer's individual and collective comments to the quality of a paper, I estimate the following regression equation:

\begin{equation}
\label{eq:regression}
\begin{split}
\ln Q_{pajt}=\beta_{1} \ln I_{pajt}+\beta_{2} \ln S_{pajt}+\beta_{3} \ln C_{pajt}+\beta_{4} \ln Q_{ajt}+\beta_{5} JMP_{pajt}\\+\gamma_{j}+\gamma_{t}+\epsilon_{patj}
\end{split}
\end{equation}

where $Q_{pajt}$ is the quality of paper $p$ written by author $a$ who did her doctoral studies at university $j$ and entered the job market at year $t$. $I_{pajt}$ is the number of individual comments received by paper $p$; $S_{pajt}$ and $C_{pajt}$ are the number of seminars and conferences at which paper $p$ was presented, respectively. $Q_{ajt}$ is the quality of the author and $JMP_{pajt}$ is an indicator variable that turns 1 if the paper was a job market paper. $\epsilon_{patj}$ is the disturbance term.

Since some economics departments may have more social ties with journal editors than others \citep{colussi2018socialties}, I control for the economics department at which the candidate did her PhD ($\gamma_{j}$). Due to the time elapsing between submitting a paper and being accepted for publication at a journal, papers from ''younger'' PhD candidate cohorts are less likely to be included in the estimation sample. This may create a sample selection problem in the dependent variable. To address this problem, I introduce cohort fixed effects ($\gamma_{t}$). They also control for other cohort-specific factors that may affect the probability of publishing in a high-quality journal, such as the quality of other PhD candidates that entered the job market in the same year, or the number of PhD candidates that decided to pursue an academic career. 

Table~\ref{tab:count} presents the estimates for the impact of the number of comments given by peers' individually and collectively at research seminars and conferences.\footnote{Since the number of individual comments, seminars, and conferences enter in logs in Equation~\eqref{eq:regression}, I add 1 to the number of comments, seminars, and conferences variables in order to keep the observations with zero values in the estimation sample. Results are robust to using an inverse hyperbolic sine transformation of the variables \citep{bellamare2019inversehyperbolicsine}.} I cluster standard errors at the author level. First, I estimate Equation~\eqref{eq:regression} with the number of individual comments variable only (column (1) of Table~\ref{tab:count}). This estimation uses the full sample of publications: 806. As expected, the \textit{$\ln$ Comment} coefficient is positive and very precisely estimated. This result indicates that receiving more individual comments is positively correlated with publishing in a high-ranked journal. For example, a one standard deviation increase in the number of comments (7 comments), for a paper that received the average number of comments (10 comments), increases the quality of the journal in which the paper is published by 29\% [((ln(17)-ln(10))*0.553]. This increase would lift a paper published in a journal located in the 2nd quartile of the SJR Economics and Econometrics category (e.g., CESifo Economics Studies; SJR score: 0.851) to a journal located in the 1st quartile (e.g., Journal of Industrial Economics, SJR score: 1.059).

%\FloatBarrier

\begin{table}[tbp]
	\begin{center}
		\footnotesize
		\caption{Contribution of individual comments, seminars, and conferences to the quality of a paper}
		\label{tab:count}
		{
\def\sym#1{\ifmmode^{#1}\else\(^{#1}\)\fi}
\begin{tabular}{l*{7}{c}}
\hline\hline
                    &\multicolumn{1}{c}{(1)}       &\multicolumn{1}{c}{(2)}       &\multicolumn{1}{c}{(3)}       &\multicolumn{1}{c}{(4)}       &\multicolumn{1}{c}{(5)}       &\multicolumn{1}{c}{(6)}       &\multicolumn{1}{c}{(7)}       \\
\hline
ln Comment          &       0.553\sym{a}&                   &                   &       0.397\sym{a}&       0.380\sym{a}&       0.348\sym{a}&       0.305\sym{b}\\
                    &     (0.052)       &                   &                   &     (0.059)       &     (0.060)       &     (0.057)       &     (0.119)       \\
[1em]
ln Seminar          &                   &       0.428\sym{a}&                   &       0.288\sym{a}&       0.267\sym{a}&       0.156\sym{a}&       0.317\sym{a}\\
                    &                   &     (0.044)       &                   &     (0.050)       &     (0.048)       &     (0.047)       &     (0.112)       \\
[1em]
ln Conference       &                   &                   &       0.337\sym{a}&      -0.055       &      -0.049       &      -0.025       &      -0.250       \\
                    &                   &                   &     (0.069)       &     (0.070)       &     (0.068)       &     (0.065)       &     (0.151)       \\
[1em]
ln Author(s) quality&                   &                   &                   &                   &       0.093\sym{a}&       0.104\sym{a}&                   \\
                    &                   &                   &                   &                   &     (0.022)       &     (0.021)       &                   \\
[1em]
Job market paper    &                   &                   &                   &                   &                   &       0.494\sym{a}&       0.277\sym{c}\\
                    &                   &                   &                   &                   &                   &     (0.076)       &     (0.141)       \\
\hline
Observations        &         806       &         685       &         685       &         685       &         685       &         685       &         276       \\
R-square            &       0.376       &       0.345       &       0.258       &       0.400       &       0.421       &       0.456       &       0.282       \\
Author(s) FE        &          No       &          No       &          No       &          No       &          No       &          No       &         Yes       \\
\hline\hline
\end{tabular}
}

		\caption*{\begin{footnotesize}Note: The dependent variable is the journal's log impact factor. Estimations in Columns (1) to (6) include cohort and PhD institution fixed effects (not reported). Standard errors clustered at the author level are in parentheses. a, b, c: statistically significant at 1\%, 5\%, and 10\%, respectively. 				
		\end{footnotesize}}
	\end{center}
\end{table}

In column~(2) the number of seminars is the only independent variable. Note that the number of observations is lower than in column~(1) since, as mentioned above, there are some papers that do not provide a valid list of seminars. As expected, presenting the paper at research seminars is positively correlated with publishing the paper at a high-ranked journal. For example, a one standard deviation increase in the number of presentations (5 seminars) for a paper that was presented at the average number of research seminars (3 seminars), increases the quality of the journal in which the paper is published by 42\% [((ln(8)-ln(3))*0.428]. There is a positive correlation between the number of conferences in which a paper was presented and quality of the journal in which the paper was published (column~(3)). Specifically, a one standard deviation increase in the number of conferences (1 conference), for a paper that was presented at the average number of conferences (1 conference), increases the quality of the journal by 23\% [((ln(2)-ln(1))*0.337].

Column~(4) presents the results when the specification includes all peers' contribution variables: individual comments, research seminars, and conferences. The \textit{$\ln$ Comment} and the \textit{$\ln$ Seminar} coefficients remain positive and very precisely estimated. However, both coefficients have a lower point value than in previous estimations. This result indicates that there is a positive correlation between the number of individual comments received by a paper and number of seminars and conferences at which it is presented. Interestingly, the conference coefficient is close to zero. This result indicates that the positive association between the number of conferences in which a paper is presented and quality of the journal in which it is published disappears once I control for the number of individual comments received by a paper and seminars in which it was presented. According to the coefficients reported in column~(4), a one standard deviation increase in the number of individual comments and research seminars, for a paper that has an average number of comments and seminars, increases the quality of the journal in which the paper is published by 49\% [((ln(17)-ln(10))*0.397 + (ln(8)-ln(3))*0.288].

In column~(5), I introduce the quality of the author as an additional regressor. As expected, the quality of the author is positively correlated with quality of the journal in which the paper is published. There is also a reduction in the \textit{$\ln$ Comment} and \textit{$\ln$ Seminar} coefficients' point estimates, suggesting that these coefficients were partially capturing the positive correlation between the quality of the author and journal.\footnote{I also analyzed whether papers with more than one author had a larger quality than solo papers. The coefficient for multi-authored papers was imprecisely estimated.} Column~(6) presents the results when I control for the quality of the research idea. The job market paper coefficient is positive and very precisely estimated. According to the coefficient reported in column~(6), the quality of journals in which job market papers were published was, on average, 63\% higher than the quality of the journals in which the rest of projects were published (exp .494). The \textit{$\ln$ Comment} and \textit{$\ln$ Seminar} coefficients remain positive and precisely estimated. However, their point values, specially for \textit{$\ln$ Seminar}, are lower than in column~(5). This is consistent with the argument that scholars choose to present their most promising projects when they are invited to give a research seminar. Even when I control for the quality of the author and research idea, a one standard deviation increase in the number of comments and seminars, for a paper with average values of these variables, still increases the quality of the journal in which the paper is published by 34\% [((ln(17)-ln(10))*0.348 + (ln(8)-ln(3))*0.156].

Finally, column~(7) reports the estimations when the regression equation includes author fixed effects. This estimation controls for all variables that are author specific, such as the capacity to transform new ideas into high-quality publications, or the "contribution-threshold" each author establishes to determine whether or not a peer is included in the acknowledgments section, or author's willingness to present at research seminars and conferences.\footnote{The quality of the author, and the PhD institution and cohort fixed effects are removed from the regression equation since they are collinear with author fixed effects.} In this specification, I identify peers' contribution to the quality of a paper with the variation in the number of individual and collective comments among papers written by the same author, who were devised and began to be developed during the same period, and whose initial quality was identified by the author. Although I do not have a natural experiment that generates a random variation in the number of individual and collective comments received by a paper, I argue that, conditional on author fixed effects and the initial quality of the paper, the variation is mostly random. This enables me to lean towards a causal interpretation of estimates.

The sample in column~(7) only includes scholars that published more than one of the projects that were initiated when they were doing their doctoral studies. This leads to a large reduction in the number of observations. Despite this drop, and the increase in standard errors, the \textit{$\ln$ Comment} and \textit{$\ln$ Seminar} coefficients remain positive and precisely estimated. This result confirms that peers' individual and collective comments improve the quality of a paper. A one standard deviation increase in the number of comments, for a paper that received the average number of comments, increases the quality of the journal in which the paper is published by 16\% [((ln(17)-ln(10))*0.305]; and a one standard deviation increase in the number of seminars, for a paper that was presented in the average number of seminars, increases the quality of the journal by 31\% [((ln(8)-ln(3))*0.317]. For example, the combined effect of these increases, 47\%, would lift a paper published in Review of Economics and Statistics (8.363) to the quality level of The American Review (11.889). Presenting at conferences does not raise the quality of the journal in which the paper is published.

\begin{table}[tbp]
	\begin{center}
		\small
		\caption{Peers' contribution by quality}
		\label{tab:top10weighted}
		{
\def\sym#1{\ifmmode^{#1}\else\(^{#1}\)\fi}
\begin{tabular}{l*{7}{c}}
\hline\hline
                    &\multicolumn{1}{c}{(1)}       &\multicolumn{1}{c}{(2)}       &\multicolumn{1}{c}{(3)}       &\multicolumn{1}{c}{(4)}       &\multicolumn{1}{c}{(5)}       &\multicolumn{1}{c}{(6)}       &\multicolumn{1}{c}{(7)}       \\
\hline
ln Comment top 10   &       0.628\sym{a}&                   &                   &       0.486\sym{a}&       0.439\sym{a}&       0.401\sym{a}&       0.246\sym{c}\\
                    &     (0.067)       &                   &                   &     (0.070)       &     (0.070)       &     (0.068)       &     (0.132)       \\
[1em]
ln Comment rest     &       0.094       &                   &                   &       0.074       &       0.100\sym{c}&       0.090       &       0.202       \\
                    &     (0.057)       &                   &                   &     (0.061)       &     (0.061)       &     (0.058)       &     (0.131)       \\
[1em]
ln Seminar top 10   &                   &       0.477\sym{a}&                   &       0.304\sym{a}&       0.282\sym{a}&       0.191\sym{a}&       0.305\sym{b}\\
                    &                   &     (0.050)       &                   &     (0.050)       &     (0.049)       &     (0.049)       &     (0.139)       \\
[1em]
ln Seminar rest     &                   &      -0.035       &                   &      -0.080       &      -0.061       &      -0.134\sym{c}&       0.015       \\
                    &                   &     (0.081)       &                   &     (0.078)       &     (0.076)       &     (0.074)       &     (0.161)       \\
[1em]
ln Conference top   &                   &                   &       0.587\sym{a}&       0.388\sym{b}&       0.349\sym{b}&       0.470\sym{a}&       0.262       \\
                    &                   &                   &     (0.202)       &     (0.168)       &     (0.165)       &     (0.174)       &     (0.418)       \\
[1em]
ln Conference rest  &                   &                   &       0.273\sym{a}&      -0.095       &      -0.088       &      -0.077       &      -0.298\sym{c}\\
                    &                   &                   &     (0.073)       &     (0.072)       &     (0.070)       &     (0.067)       &     (0.154)       \\
[1em]
ln Author(s) quality&                   &                   &                   &                   &       0.073\sym{a}&       0.085\sym{a}&                   \\
                    &                   &                   &                   &                   &     (0.021)       &     (0.021)       &                   \\
[1em]
Job market paper    &                   &                   &                   &                   &                   &       0.489\sym{a}&       0.265\sym{c}\\
                    &                   &                   &                   &                   &                   &     (0.074)       &     (0.141)       \\
\hline
Observations        &         806       &         685       &         685       &         685       &         685       &         685       &         276       \\
R-square            &       0.410       &       0.352       &       0.260       &       0.435       &       0.447       &       0.481       &       0.301       \\
Author(s) FE        &          No       &          No       &          No       &          No       &          No       &          No       &         Yes       \\
\hline\hline
\end{tabular}
}

		\caption*{\begin{footnotesize}Note: The dependent variable is the journal's log impact factor. Estimations in Columns (1) to (6) include cohort and PhD institution fixed effects (not reported). Standard errors clustered at the author level are in parentheses. a, b, c: statistically significant at 1\%, 5\%, and 10\%, respectively.
		\end{footnotesize}}
	\end{center}
\end{table}

In previous estimations, I assumed that all individual and collective comments contributed equally to increase the quality of a paper. However, it seems reasonable to expect that individual comments from top scholars, or comments received at presentations at top economics departments, or leading conferences, contribute more to improve the quality of a research project.

Table~\ref{tab:top10weighted} presents the results when individual and collective comments are distinguished by quality. Column~(1) shows that comments given by top 10 scholars have a much larger positive correlation with quality of the paper than comments offered by other scholars. Column~(2) reports that presenting the paper at a top 10 economics department has a strong positive association with publishing the paper at a high-ranked journal. However, presenting the paper at a non-top 10 economics department has no correlation with quality of the paper. Presenting the paper at a major economics conference (American Economic Association, European Economic Association, and the Royal Economic Society) has a strong positive correlation with quality of the journal in which the paper is published. Presenting at other conferences also has a positive coefficient, although its point value is lower. The quality of the author (column~(5)) and the job market status of the paper (column~(6)) increase the quality of the journal in which the paper is published. In these specifications, the comments offered by top scholars, giving a seminar at top departments, or presenting the paper at a leading conference have a stronger positive association with quality of the paper than comments by non-top scholars or presenting the paper at non-top departments or conferences. When I control for author fixed effects (column~(7)), comments given by top 10 scholars have a larger positive impact on quality of the paper than comments given by non-top scholars. However, the difference between the coefficients is not large. I find that presenting the paper at a top 10 economics department has a positive effect on the quality of the paper. However, presenting the paper in a non-top economics department has no effect on the quality of the journal in which the paper is published. The coefficient for top conferences is positive, but imprecisely estimated. Surprisingly, presenting at a non-top conference has a negative effect on the quality of the journal in which the paper is published.\footnote{I also ran regressions using top 5 as the quality threshold for scholars and seminars. Results, not reported, are qualitatively and quantitatively similar to those presented in Table~\ref{tab:top10weighted}.}

As explained above, there is an important number of publications (121 out of 806) that acknowledged the comments received by participants at research seminars and conferences, but did not list the institutions at which these seminars were hold, or the name of the conferences. To test the robustness of my results, I re-estimate all specifications with the whole sample (806 observations instead of 685) and removing the number of seminars and conferences variables from the regression equation. The estimates for the \textit{$\ln$ Comment} coefficient should be taken with caution. Since the number of individual comments is correlated with the number of seminars and conferences, the \textit{$\ln$ Comment} coefficient may also capture the effect that seminars and conferences have on the quality of the journal in which a paper is published. Table~\ref{tab:onlycomments_top10} in the Appendix confirms that individual comments have a strong positive effect on the quality of the journal in which the paper is published (columns (1) to (4)). Estimates also confirm that the individual comments given by top scholars have a stronger effect on the quality of the journal in which the paper is published than comments provided by non-top scholars (columns~(5) to~(8)).

Finally, I analyze whether some type of papers benefit more from individual and collective comments than others. Following the methodology used in \cite{card2020refereesgenderneutral}, I classify papers as theoretical, empirical, structural, or experimental based on the counting of some specific words.\footnote{The words used to identify a theoretical paper are proposition, theorem, lemma, proof, model, and theory; for an empirical paper: data, standard error, table, regression, difference-in-differences, and empirical; for a structural paper: structural, BLP, maximum likelihood, mixture, simulation, and calibration; and for an experimental paper: field experiment, RCT, laboratory, subjects, and survey. The category with the largest number of words determines the type the paper belongs to.} 61\% of papers in the sample are experimental, and 36\% are theoretical. I select these categories and compare whether peers' comments have a larger impact on empirical than on theoretical papers. I expand Equation~\eqref{eq:regression} with a dummy variable that turns one if the paper is empirical, and interact the comment, seminar, and conference coefficients with the empirical dummy variable. 

Table~\ref{tab:empirical} in the Appendix presents the results. On average, empirical and theoretical papers are published in journals that have a similar impact factor. Individual comments and research seminars have a similar effect on theoretical and empirical papers. The results for conferences are not robust. Columns~(3) and~(4) presents the results when peer effects are distinguished by quality. Comments from high-quality scholars have similar effects on theoretical and empirical papers, whereas comments from non-top scholars have a larger positive effect on empirical papers. Research seminars at top economics departments have similar effects on theoretical and empirical papers, whereas seminars at non-top departments have a larger positive impact on empirical papers. Results for conferences are not robust.

\section{Conclusions}
\label{sec:conclusions}

A scholar's knowledge is limited and, therefore, is unaware of all the elements that may contribute to improve the quality of her research. To discover these elements, she relies on peers, who at research seminars, conferences, or through conversations, identify limitations in the research project and suggest avenues to improve it. In this paper, I measured how much these comments and suggestions improve the quality of research. Since the number of suggestions a paper receives is not independent from the quality of the research idea and author, I use a sample of papers that enables me to control for these variables: the research projects of job market candidates in economics. I find that a one standard deviation increase in the number of individual comments and research seminars increases the quality of the journal in which the paper is published by 47\%. I find that comments provided by top scholars have a stronger positive effect on the quality of the paper than comments given by non-top scholars. I also show that while presenting a paper at top economics departments has a strong positive effect on the quality of the paper, presenting at non-top economics departments has no effect. I find that presenting at conferences, even at the top ones, is not associated with publishing in a high-ranked journal, once I control for the number of individual comments and seminars. Peers' comments have similar effect on theoretical and empirical papers.

My results confirm that peers' individual and collective comments have a large positive effect on the quality of research projects, specially when they come from top scholars or are received when presenting the paper at a top economics department. From a policy perspective, these results justify the use of public funding to organize research seminars, interact with other scholars, and finance stays at top economics departments.

\clearpage

\appendix \label{app:all}

\setcounter{equation}{0}
\renewcommand\theequation{A.\arabic{equation}}

\setcounter{figure}{0}

\renewcommand\thefigure{A.\arabic{figure}}

\setcounter{table}{0}

\renewcommand\thetable{A.\arabic{table}}

{\begin{table}[tbp] \centering
\newcolumntype{C}{>{\centering\arraybackslash}X}

\caption{Graduate programs in economics included in the sample}
\label{tab:phd_programs}
{\footnotesize
\begin{tabularx}{\textwidth}{lCrrr}

\toprule
{University}&{Cohort}&{Graduates}&{Publications} \tabularnewline
\midrule\addlinespace[1.5ex]
Arizona State&2016, 2018&12&0 \tabularnewline
Boston&2014, 2015, 2016, 2017, 2018&84&11 \tabularnewline
Boston College&2018&5&0 \tabularnewline
Brown&2014, 2015, 2016, 2018&30&12 \tabularnewline
Chicago&2008, 2010, 2011, 2013, 2015, 2018&93&40 \tabularnewline
Columbia&2017, 2018&39&4 \tabularnewline
Cornell&2017, 2018&38&3 \tabularnewline
Duke&2016, 2017, 2018&31&8 \tabularnewline
George Washington&2013, 2014, 2015, 2016, 2017, 2018&39&10 \tabularnewline
Georgetown&2015, 2016, 2017, 2018&18&4 \tabularnewline
Harvard&2011, 2012, 2013, 2014, 2015, 2016, 2017, 2018&196&76 \tabularnewline
Iowa State&2017&3&0 \tabularnewline
Johns Hopkins&2011, 2012, 2015, 2016, 2017&34&13 \tabularnewline
MIT&2000, 2001, 2002, 2003, 2004, 2005, 2006, 2007, 2008, 2010, 2011, 2012, 2016 , 2017, 2018&242&192 \tabularnewline
Maryland&2015, 2016, 2017, 2018&45&2 \tabularnewline
Michigan&2018&17&0 \tabularnewline
Michigan State&2010, 2011, 2012, 2013, 2014, 2015, 2016, 2017, 2018&84&35 \tabularnewline
Minnesota&2016, 2017, 2018&38&1 \tabularnewline
New York&2017, 2018&33&0 \tabularnewline
Northwestern&2000, 2001, 2003, 2004, 2005, 2006, 2007, 2008&78&61 \tabularnewline
Notre Dame&2011, 2012, 2013, 2015, 2016, 2018&20&12 \tabularnewline
Ohio State&2018&10&4 \tabularnewline
Oregon&2007, 2008, 2010, 2011, 2012, 2013, 2015, 2016, 2018&31&15 \tabularnewline
Penn State&2017&8&0 \tabularnewline
Pittsburgh&2014, 2015, 2016, 2017, 2018&23&13 \tabularnewline
Princeton&2014, 2015, 2016, 2017, 2018&94&16 \tabularnewline
Rutgers&2010, 2011, 2012, 2013, 2014, 2015, 2016, 2017, 2018&47&8 \tabularnewline
Southern California&2016, 2017, 2018&23&7 \tabularnewline
Stanford&2008, 2011, 2012, 2014, 2015, 2016, 2017, 2018, 2019&129&63 \tabularnewline
Texas Austin&2015, 2016, 2018&30&4 \tabularnewline
UC Berkeley&2000, 2002, 2003, 2004, 2005, 2006, 2007, 2009, 2011, 2018&187&128 \tabularnewline
UC Davis&2018&7&1 \tabularnewline
UC Irvine&2014, 2015, 2016, 2018&45&31 \tabularnewline
UC Los Angeles&2017, 2018&33&3 \tabularnewline
UC San Diego&2016, 2017, 2018&46&14 \tabularnewline
UC Santa Barbara&2016&12&6 \tabularnewline
UC Santa Cruz&2014, 2015, 2016, 2017, 2018&29&3 \tabularnewline
Vanderbilt&2014, 2015, 2016, 2018&22&9 \tabularnewline
Virginia&2000, 2001, 2002, 2003, 2004, 2005, 2006, 2007, 2009, 2011, 2016, 2017, 2018&55&14 \tabularnewline
Wisconsin-Madison&2018&16&0 \tabularnewline
Yale&2015, 2016, 2018&41&9 \tabularnewline
\bottomrule \addlinespace[1.5ex]

\end{tabularx}
\begin{flushleft}
\footnotesize Note: Cohort is the year when job market candidates were announced.
\end{flushleft}
}
\end{table}
}

\begin{table}[tbp] \centering
\newcolumntype{C}{>{\centering\arraybackslash}X}

\caption{Information about the sample}
\label{tab:data_collection}
\begin{tabularx}{12.5cm}{lr}

\toprule
Job market candidates&2067 \tabularnewline
Potential papers&5118 \tabularnewline
Job market candidates with a publication in a SJR journal&551 \tabularnewline
Publications in a SJR journal&806 \tabularnewline
Main estimation sample&685 \tabularnewline
\bottomrule \addlinespace[1.5ex]

\end{tabularx}
\end{table}

%\iffalse

\begin{table}[htbp]
	\begin{center}
		\footnotesize
		\caption{Contribution of peers' individual comments to the quality of a paper. Full sample}
		\label{tab:onlycomments_top10}
		{
\def\sym#1{\ifmmode^{#1}\else\(^{#1}\)\fi}
\begin{tabular}{l*{8}{c}}
\hline\hline
                    &\multicolumn{1}{c}{(1)}       &\multicolumn{1}{c}{(2)}       &\multicolumn{1}{c}{(3)}       &\multicolumn{1}{c}{(4)}       &\multicolumn{1}{c}{(5)}       &\multicolumn{1}{c}{(6)}       &\multicolumn{1}{c}{(7)}       &\multicolumn{1}{c}{(8)}       \\
\hline
ln Comment          &       0.553\sym{a}&       0.524\sym{a}&       0.425\sym{a}&       0.466\sym{a}&                   &                   &                   &                   \\
                    &     (0.052)       &     (0.052)       &     (0.051)       &     (0.115)       &                   &                   &                   &                   \\
[1em]
ln Comment top 10   &                   &                   &                   &                   &       0.628\sym{a}&       0.564\sym{a}&       0.470\sym{a}&       0.373\sym{a}\\
                    &                   &                   &                   &                   &     (0.067)       &     (0.066)       &     (0.064)       &     (0.133)       \\
[1em]
ln Comment rest     &                   &                   &                   &                   &       0.094       &       0.124\sym{b}&       0.094\sym{c}&       0.222\sym{c}\\
                    &                   &                   &                   &                   &     (0.057)       &     (0.056)       &     (0.053)       &     (0.116)       \\
[1em]
ln Author(s) quality&                   &       0.104\sym{a}&       0.110\sym{a}&                   &                   &       0.087\sym{a}&       0.095\sym{a}&                   \\
                    &                   &     (0.020)       &     (0.019)       &                   &                   &     (0.019)       &     (0.019)       &                   \\
[1em]
Job market paper    &                   &                   &       0.563\sym{a}&       0.455\sym{a}&                   &                   &       0.518\sym{a}&       0.413\sym{a}\\
                    &                   &                   &     (0.067)       &     (0.100)       &                   &                   &     (0.064)       &     (0.099)       \\
\hline
Observations        &         806       &         806       &         806       &         341       &         806       &         806       &         806       &         341       \\
R-square            &       0.376       &       0.403       &       0.457       &       0.253       &       0.410       &       0.428       &       0.473       &       0.266       \\
Author(s) FE        &          No       &          No       &          No       &         Yes       &          No       &          No       &          No       &         Yes       \\
\hline\hline
\end{tabular}
}

		\caption*{\begin{footnotesize}Note: The dependent variable is the journal's (log) impact factor. Estimations in Columns (1) to (3), and (5) to (7), include cohort and PhD institution fixed effects (not reported). Standard errors clustered at the author level are in parentheses. a, b, c: statistically significant at 1\%, 5\%, and 10\%, respectively. 
		\end{footnotesize}}
	\end{center}
\end{table}

\begin{table}[htbp]
	\begin{center}
		\scriptsize
		\caption{Theoretical vs. empirical papers}
		\label{tab:empirical}
		{
\def\sym#1{\ifmmode^{#1}\else\(^{#1}\)\fi}
\begin{tabular}{l*{4}{c}}
\hline\hline
                    &\multicolumn{1}{c}{(1)}       &\multicolumn{1}{c}{(2)}       &\multicolumn{1}{c}{(3)}       &\multicolumn{1}{c}{(4)}       \\
\hline
Empirical           &      -0.076       &       0.194       &      -0.038       &       0.286       \\
                    &     (0.198)       &     (0.402)       &     (0.180)       &     (0.346)       \\
[1em]
ln Comment          &       0.279\sym{a}&       0.215       &                   &                   \\
                    &     (0.088)       &     (0.187)       &                   &                   \\
[1em]
ln Comment*Empirical&       0.112       &       0.076       &                   &                   \\
                    &     (0.096)       &     (0.203)       &                   &                   \\
[1em]
ln Seminar          &       0.224\sym{a}&       0.282\sym{b}&                   &                   \\
                    &     (0.078)       &     (0.140)       &                   &                   \\
[1em]
ln Seminar*Empirical&      -0.104       &       0.058       &                   &                   \\
                    &     (0.092)       &     (0.185)       &                   &                   \\
[1em]
ln Conference>0     &      -0.183       &       0.083       &                   &                   \\
                    &     (0.121)       &     (0.205)       &                   &                   \\
[1em]
ln Conference*Empirical&       0.257\sym{c}&      -0.465\sym{c}&                   &                   \\
                    &     (0.140)       &     (0.267)       &                   &                   \\
[1em]
ln Comment top 10   &                   &                   &       0.425\sym{a}&       0.506\sym{a}\\
                    &                   &                   &     (0.115)       &     (0.170)       \\
[1em]
ln Comment top 10*Empirical&                   &                   &      -0.012       &      -0.356       \\
                    &                   &                   &     (0.129)       &     (0.221)       \\
[1em]
ln Comment rest     &                   &                   &      -0.022       &      -0.143       \\
                    &                   &                   &     (0.098)       &     (0.193)       \\
[1em]
ln Comment rest*Empirical&                   &                   &       0.143       &       0.461\sym{b}\\
                    &                   &                   &     (0.115)       &     (0.202)       \\
[1em]
ln Seminar top 10   &                   &                   &       0.244\sym{a}&       0.421\sym{b}\\
                    &                   &                   &     (0.082)       &     (0.161)       \\
[1em]
ln Seminar top 10*Empirical&                   &                   &      -0.088       &      -0.206       \\
                    &                   &                   &     (0.098)       &     (0.222)       \\
[1em]
ln Seminar rest     &                   &                   &      -0.181       &      -0.456       \\
                    &                   &                   &     (0.152)       &     (0.357)       \\
[1em]
ln Seminar rest*Empirical&                   &                   &       0.067       &       0.781\sym{c}\\
                    &                   &                   &     (0.175)       &     (0.427)       \\
[1em]
ln Conference top   &                   &                   &       0.718\sym{b}&      -0.293       \\
                    &                   &                   &     (0.353)       &     (0.794)       \\
[1em]
ln Conference top*Empirical&                   &                   &      -0.396       &       1.337       \\
                    &                   &                   &     (0.391)       &     (0.918)       \\
[1em]
ln Conference rest  &                   &                   &      -0.222\sym{c}&       0.025       \\
                    &                   &                   &     (0.122)       &     (0.206)       \\
[1em]
ln Conference rest*Empirical&                   &                   &       0.264\sym{c}&      -0.713\sym{a}\\
                    &                   &                   &     (0.147)       &     (0.267)       \\
[1em]
ln Author(s) quality&       0.101\sym{a}&                   &       0.079\sym{a}&                   \\
                    &     (0.021)       &                   &     (0.021)       &                   \\
[1em]
Job market paper    &       0.504\sym{a}&       0.274\sym{c}&       0.506\sym{a}&       0.277\sym{c}\\
                    &     (0.078)       &     (0.154)       &     (0.076)       &     (0.147)       \\
\hline
Observations        &         666       &         264       &         666       &         264       \\
R-square            &       0.459       &       0.312       &       0.489       &       0.386       \\
Author(s) FE        &          No       &         Yes       &          No       &         Yes       \\
\hline\hline
\end{tabular}
}

		\caption*{\begin{footnotesize}Note: The dependent variable is the journal's log impact factor. Estimations in Columns (1) and (3) include cohort and PhD institution fixed effects (not reported). Standard errors clustered at the author level are in parentheses. a, b, c: statistically significant at 1\%, 5\%, and 10\%, respectively.
		\end{footnotesize}}
	\end{center}
\end{table}

\end{document}